\documentclass[10pt]{iopart}
\usepackage{harvard}
\begin{document}

\title{Adding an energy-like conservation law to the leapfrog
  integrator} \author{A. C. Maggs,} \address{Laboratoire PCT,
  Gulliver CNRS-ESPCI UMR 7083, 10 rue Vauquelin, 75231 Paris Cedex
  05.}
\begin{abstract}
  The leapfrog integrator is widely used because of its excellent
  stability in molecular dynamics simulation. This is recognized as
  being due to the existence of a discrete variational structure of
  the equations.  We introduce a modified leapfrog method which
  includes an additional energy-like conservation law by embedding a
  molecular dynamics simulation within a larger dynamical system.
\end{abstract}
\section{Introduction}

The leapfrog integrator for molecular dynamics is known to display
several exceptional features which allow it to have superior long-time
dynamic stability compared with many higher-order integrators. This
exceptional nature is due to the exact conservation of momentum and
symplectic structure by the discretization \cite{geometric}. The
symplectic nature is particularly important for statistical mechanics
applications because it tells us that there is a well defined density
in phase space which is conserved as in the classic Liouville theorem;
this density is the basis for the construction of the Gibbsian
approach to statistical mechanics \cite{gibbs}. In addition the
existence of a backward error analysis \cite{reich} tells us that we
are observing trajectories which are those of a perturbed Hamiltonian
which is close to that which we are interested in.

Despite these numerous advantages there is one important quantity
which is not exactly conserved, this is the energy. It fluctuates on
small time scales and can drift in the very longest simulations.This
is normally countered by introducing a coupling to a thermostat
\cite{nose,hoover,frenkel}, but this leads to a change of ensemble
from micro-canonical to canonical.  The aim of the present paper is
the construction of an integrator that is similar in many ways to the
leapfrog method, but which is embedded in a larger dynamic system. The
dynamics of the larger system are such that the original energy
emerges as an additional conservation law.

There are many trivial (and bad) manners to impose such an energy
conservation- for instance one can regularly rescale the particle
velocities. However, such arbitrary modifications to the dynamics
break the symplectic structure which is a disaster for applications in
statistical mechanics. Our approach to build a larger dynamic set of
equations via variational methods in such a way that we can construct
the discrete Hamiltonian of the system and thus explicitly understand
the phase-space structure of the extended dynamical system.

This extended dynamical system is built using several components. We
start by considering a discretized Lagrangian, in which the extra
conservation is imposed by a Lagrange multiplier. From this dynamic
system we build a discretized Hamiltonian. This Hamiltonian has a
common defect that occurs in constrained systems (such as
electrodynamics) -- there is no momentum which corresponds to the
multiplier. The solution is to add additional terms to the Hamiltonian
which are zero, but which nevertheless generate independent dynamical
equations for the multiplier. The logic is very close to the treatment
of the potential in electrodynamics which has the formal role of being
the Lagrange multiplier in imposing Gauss' law \cite{dirac}.

\section{Variational integrators}

We firstly resume how to pass from a discretized Lagrangian to the
leapfrog integrator before generalizing to our more complicated
constrained system: Newtons equations of motion for particles moving under velocity
independent forces can be found by considering the variational problem
\begin{equation}
  \delta \int \left  \{  \frac{m}{2} \left (\frac{d q}{dt} \right )^2 - V(q)
  \right  \} dt =0
\end{equation}
In the following we will take all masses to be identical, and will
allow $q$ to denote a $N\times d$ dimensional vector corresponding to
$N$ particles moving in $d$ dimensional space. This Lagrangian can be
discretized by replacing derivatives by finite differences evaluated
every $\tau$ so that $t_k=k\tau$:
\begin{equation}
  L_k= m \frac {( q_{k+1} - q_k)^2}{2 \tau^2} -V(q_k) \label{eq:L}
\end{equation}
The discretized action principle is then \cite{variation,marsden}
\begin{equation}
  \delta   \sum_k  \tau L_k = 0
\end{equation}
This variation then gives simple partial derivatives with respect to
$x_k$ so that
\begin{equation}
  m (  q_{k+1} + q_{k-1} -2 q_k ) + V'(q_k)=0 \label{eq:stepping}
\end{equation}
which is indeed a version of the leapfrog algorithm \cite{frenkel}. In
the continuous time limit the energy is exactly conserved.
\begin{equation}
  U =    \frac{m}{2} \left (\frac{d q}{dt} \right )^2 + V(q)  \label{eq:U}
\end{equation}
However any time stepping procedure such as eq.~(\ref{eq:stepping})
leads to a breakdown in the conservation of energy. The first step of
our modified procedure will be to take the energy eq.~(\ref{eq:U}) and
add it as a constraint to the original Lagrangian density
eq.~(\ref{eq:L}).
\begin{eqnarray}
  L_k= &m \frac {( q_{k+1} - q_k)^2}{2 \tau^2} -V(q_k) + \nonumber \\
  &\lambda_k \left ( m \frac {( q_{k+1} - q_k)^2}{2 \tau^2} + V(q_k)
    -U \right ) \label{eq:quasi}
\end{eqnarray}
We will call the second line of eq.~(\ref{eq:quasi}) the {\it
  quasi-energy}. In the following the dynamic schemes that we propose
will conserve this discretized quasi-energy to machine accuracy.  In
this expression $\lambda_k$ is a Lagrange multiplier whose dynamics
will be developed in the following sections.  In the continuous time
limit clearly $\lambda=0$.

The main questions that will arrise are the following: In the presence
of the extended dynamical system eq.~(\ref{eq:quasi}) how do we
interpret Liouville's theorem?  What are the corresponding momentum
variables for the discretized evolution equations that comes from
eq.~(\ref{eq:quasi}).  In order to answer these questions we pass from
the Lagrangian description to a Hamiltonian form for the dynamics.

\section{Discrete Hamiltonians}
We will need to introduce a slightly more formal notation in order to
pass from the above Lagrangian formulation to a discrete Hamiltonian
form. However this notation is such that we find expressions which are
very close to those in standard treatments of Hamiltonian dynamics.
We firstly introduce the finite time difference operator
\begin{equation}
  \Delta q_k = (q_{k+1} -q_k)/\tau
\end{equation}
We also need its adjoint $\Delta^*$ which is defined so that
\begin{equation}
  \sum a_k \Delta q_k= -\sum q_k (a_k -a_{k-1})/\tau = \sum a_k
  \Delta^* q_k
\end{equation}
Thus we see that
\begin{equation}
  \Delta^* q_k = - \Delta q_{k-1}
\end{equation}
There is a natural shift of unity in indices when performing the
discrete analogy of integrating by parts.

The Lagrangian equations of motion are then
\begin{equation}
  \Delta^* \frac{\partial L}{\partial \Delta q_k} + \frac{\partial L}{\partial q_k}=0
\end{equation}
which is very close to their form in the continuum limit.  We now
define the momentum variables:
\begin{equation}
  p_{k+1} = \frac{\partial L}{\partial \Delta q_k} = m (q_{k+1} -q_k)
  (1+\lambda_k)/\tau \label{eq:p}
\end{equation}
and construct the Hamiltonian as usual
\begin{eqnarray}
  H(q_k, p_{k+1}) &=  p_{k+1} \Delta q_k - L \nonumber \\
  &= \frac{ p_{k+1}^2}{2(1+\lambda_k) m} + V_k (1- \lambda_k) +
  \lambda_k U \label{eq:H}
\end{eqnarray}

Let us consider the equations of motion which come from applying
Hamilton's principle to eq.~(\ref{eq:H}). We calculate
\begin{equation}
  \delta \sum_k \left [  p_{k+1} \Delta q_k - H(p_{k+1}, q_k) \right]
  =0 \label{eq:hamilton}
\end{equation} 
and find
\begin{eqnarray}
  \frac{\partial H}{\partial q_k} = - \Delta &p_k \label{eq:delta1}\\
  \frac{\partial H}{\partial p_{k+1}} =  \Delta& q_{k}
\end{eqnarray}
which is the discretized form of the Hamiltonian equations of motion.
More explicitly we have
\begin{eqnarray}
  \Delta p_k &=  p_{k+1}-p_k= -\tau (1-\lambda_k) V'_k \label{eq:dp}\\
  \Delta q_k &= q_{k+1}-q_k=\tau
  \frac{p_{k+1}}{m(1+\lambda_k) \label{eq:dq}}
\end{eqnarray}
When $\lambda_k=0$ this corresponds to the standard alternating update
in the leapfrog algorithm. 

Now consider the equation which comes from varying the Lagrange
multiplier $\lambda_k$:
\begin{equation}
  \frac {\partial H} {\partial \lambda_k}= U-V(q_k) -
  \frac{p_{k+1}^2}{2 m (1+\lambda_k)^2} = W_k=0
  \label{constrain}
\end{equation}
which is just the equation for  the quasi-energy in
the Hamiltonian picture. Note this quasi-energy conservation does not
imply that the Hamiltonian eq.~(\ref{eq:H}) is itself conserved,
however if they are numerically close one might hope that the
stability of the algorithm is also improved for $H$.

If $\lambda_k$ were a true dynamic degree of freedom we would have
deduced from eq.~(\ref{constrain}), in analogy to eq.~(\ref{eq:delta1}) that
\begin{equation}
-\Delta \pi_k = W_k(q_k, p_{k+1},\lambda_k)
\end{equation}
where $\pi_k$ the conjugate momentum to $\lambda_k$. We discover that in
order to have a full Hamiltonian description of the system we must add
this extra degree of freedom, but also that $\pi_k=0$ for all $k$ in
order conserve the quasi-energy. We will show this is possible later,
but firstly move on to the practical question of implementation of the
algorithm.

\section{Integration loop}
The equations eq.~(\ref{eq:dp}), eq.~(\ref{eq:dq}) together with the
constraint equation $W_k=0$, eq.~(\ref{constrain}), tell us how the positions
and momenta of the particles evolve within a time step. We now show
that the equations have explicit (non-iterative) solutions that
require only small modifications of the usual leapfrog step.

We firstly take eq.~(\ref{constrain}) and express $p_{k+1}$ in terms
of $p_k$.
\begin{equation}
  2m(U-V_k) = \frac { (p_k + \tau (1-\lambda_k) f_k )^2 }{(1+\lambda_k)^2}
\end{equation}
or
\begin{equation}
  S(1+\lambda_k)^2 = p_k^2 + 2\tau p_kf_k (1-\lambda_k) + \tau^2 f_k^2 (1-\lambda_k)^2
\end{equation}
with $f_k$ the force.  Thus
\begin{eqnarray}
  \lambda_k^2 (S - \tau^2 f_k^2) + \lambda_k (2 S + 2 \tau p_k\cdot f_k+2\tau^2
  f_k^2) + \nonumber \\(S - 2\tau p_k\cdot f_k -\tau^2 f_k^2 -p_k^2)=0 \label{eq:lambda}
\end{eqnarray}
with $S=2m (U-V_k)$. This is a simple quadratic equation for $\lambda_k$
which involves quantities which are already calculated within a
leapfrog integration loop. In practice  $\lambda$ remains small
throughout our simulations and its value 
is close to
\begin{equation}
\lambda \approx - \tau p \cdot f/S \label{eq:approx}
\end{equation}

We can integrate the equations with the following loop
\begin{quote}
  know $\lambda_{k-1}, q_{k}, p_k$\\
  \hspace{1cm } calculate $f_k( q_k) $\\
  calculate $\lambda_k $   eq.~(\ref{eq:lambda})\\
  calculate $p_{k+1} $ eq.~(\ref{eq:dp})\\
  eq.~(\ref{constrain})  satisfied at this moment of cycle\\
  calculate $q_{k+1}  $ eq.~(\ref{eq:dq})\\
  know $\lambda_{k}, q_{k+1}, p_{k+1}  $\\
\end{quote}

This is the practical generalisation of the generalised leapfrog
method with the addition of an exact conservation of the quasi-energy.

\section{Momentum for $\lambda$}

The equations of motion as stated above are adequate to implement the
algorithm. However they contain a formal weakness. While $p$ and $q$
evolution is the result of a variational principle
eq.~(\ref{eq:hamilton}) this is not true of $\lambda$. The Lagrange
multiplier is simply slaved to impose energy conservation.  Such
slaved variables are well known in quantum mechanics, indeed the
electrostatic potential is an example of such a variable; this
explains the initial difficulties in the quantisation in
electrodynamics due to the lack of an obvious conjugate momentum.  We
now show how to render the equation for the evolution of $\lambda$
autonomous, and thus better understand the phase space of the enlarged
dynamic system. We will use methods which are rather similar to those
invented in electrodynamics \cite{dirac,leimkuhler} where the
electrostatic potential also has a role which is similar to a Lagrange
multiplier. There are clear analogies too with our previous work on
local simulation algorithms for charged media \cite{mc1,mc2}. The
first problem with the equation eq.~(\ref{eq:H}) is that it does not
include a momentum variable, $\pi_k$ which is conjugate to $\lambda_k$.
We correct this deficit with the following ansatz: we add an extra
term
\begin{equation}
  \pi_{k+1} \mu_k( q_k, p_{k+1}, \lambda_k) \label{eq:mu}
\end{equation}
where $\mu_k$ is a function that we will construct later. This leads
to the following equations of motion:
\begin{eqnarray}
  \Delta \lambda_k =& \frac{\partial H}{\partial \pi_{k+1}}
  =\mu_k \label{eq:mu2}
  \\
  - \Delta \pi_k =& \frac {\partial H} {\partial \lambda_k } = W_k +
  \pi_{k+1} \frac{\partial \mu_k}{\partial \lambda_k} \label{eq:pi}
\end{eqnarray}

We now use the idea of weak constraints: For arbitrary functions
$\mu_k$ we have a Hamiltonian system.  We will show that the correct
choice of the function $\mu_k(p_{k+1}, q_k, \lambda_k) $ allows us to
impose both the conservation of eq.~(\ref{constrain}) but is also
compatible with $\Delta \pi_k=0$. We then start the dynamic system in
the state $\pi_0=0$ and this remains true for all further times in the
dynamics.

We now show that the function $\mu_k$ is indeed only a function of the
objects $(q_k, p_{k+1}, \lambda_k )$. We proceed by considering $W_k$
at two successive time steps, imposing conservation of the
quasi-energy
\begin{equation}
  W_k(q_k, p_{k+1}, \lambda_k) = W_{k+1}  (q_{k+1}, p_{k+2},
  \lambda_{k+1}) \label{eq:W}
\end{equation}
We now eliminate the variables $q_{k+1}$ and $p_{k+2} $ from the right
hand side using eq.~(\ref{eq:dq}) and eq.~(\ref{eq:dp}). If we do so
the right hand side of eq.~({\ref{eq:W}) is a function of $q_k,
  p_{k+1}, \lambda_{k+1}$ We can thus, in principle solve for
  $\lambda_{k+1}$ and write the evolution in the form
  \begin{equation}
    \lambda_{k+1} = \lambda_k + \mu_k (q_k, p_{k+1} , \lambda_k)
  \end{equation}
  as needed for the dynamics of $\pi_k$ , eq.~(\ref{eq:mu}),
  eq.~(\ref{eq:mu2})



\section{Phase space and Liouville}

We have succeeded in embedding our original system of particle
dynamics in a larger system such that the quasi-energy is exactly
conserved. To do so we were obliged to introduce two new variables
$\lambda_k$ which started as a simple Lagrange multiplier and $\pi_k$
which is the conjugate momentum. For a system of $N$ particles in
$d$-dimensional space this gives us a phase-space of dimensions
$2dN+2$.

We now study the Jacobian of the discrete evolution equations to show
that the constrained dynamical systems are compatible with the assumed
measure. The maps eq.~(\ref{eq:dp}) and eq.~(\ref{eq:dq}) are easily seen to
have unit Jacobians on the phase space defined by the variables
$(q,p,\lambda, \pi)$. The evolution equations for $\lambda$ require
slightly more study. The Jacobian for
\begin{math}
  \lambda_{k+1} = \lambda_k + \mu_k(q_k, p_{k+1}, \lambda_k)
\end{math}
is given by $J= 1 + \partial \mu_k/\partial \lambda_k$.  while we see
we can re-arrange eq.~(\ref{eq:pi}) to give
\begin{math}
  \pi_{k+1} =\pi_k/(1 + \partial \mu_k/\partial \lambda_k)
\end{math}.
It is thus the product of these two factors which ensures that the
Jacobian of a full time-step is indeed unity. We thus see that
introduction of the ``dummy'' momentum $\pi$ has absorbed the
fluctuations in phase space volumes that would otherwise result from
the use of the Lagrange multiplier.

Thus we have a complete set of Hamiltonian dynamics on the extended
phase space with the extra pair of variables $\lambda, \pi$ which are
now fully autonomous. We however {\it choose special initial
  conditions} $\pi=0$ that lead to the exact imposition of the energy
conservation.  We conclude that we have a phase space measure of the
form
\begin{math}
  dq\, dp\, d\lambda\, d\pi
\end{math}, with conservation laws imposing constant quasi-energy,
particle momentum, and $\pi$ \cite{khinchin}.

\section{Time reversal}
While we have added a conservation law to the leapfrog integrator we
have also lost a symmetry which is present in the standard leapfrog
algorithm -- it is time reversible: The more complicated quasi-energy
conserving version does not re-trace its trajectory when momenta are
reversed.  This extra symmetry can be imposed in our algorithm by
alternating the direct step (described above in section 4) with a
version in which each step in implemented in reverse order, with
$\tau\rightarrow -\tau$. The main technical difficulty is that the
equation of $\lambda_{k-1}$ given $q_k $ and $p_k $ becomes implicit
and must be solved by iteration. In practice we find that a simple
iteration procedure converges to machine precision in two steps if we
use eq.~(\ref{eq:approx}) as a starting guess for $\lambda$. The
algorithm in which a the direct and reversed step are alternated then
displays time reversal symmetry.

\section{Conclusion}

We have constructed a variational integrator which includes an
additional conserved quasi-energy. Due to the variational Hamiltonian
form we are able to study the phase space measure and understand the
discrete Liouville theorem that is implied by the
dynamics. Implementation of the algorithm requires a small overhead in
computational effort compared with the standard leapfrog
integrator. We have implemented a version of the code for the
molecular dynamics study of a truncated Lennard-Jones potential and
verified the stability of the quasi-energy during simulation. 
\vskip 1cm \bibliographystyle{jphysicsB}

\bibliography{cons}
\end{document}